# Lévy distribution in half space based on nonextensive statistical mechanics

A. K. Rajagopal† and Sumiyoshi Abe‡

†Naval Research Laboratory, Washington, DC 20375-5320

‡College of Science and Technology, Nihon University,

Funabashi, Chiba 274-8501, Japan

**Abstract**.  Probability distributions defined on the half space are known to be quite different from those in the full space. Here, a nonextensive entropic treatment is presented for the half space in an analytic and self-consistent way. In this development, the ordinary first moment of the random variable $X$ is divergent in contrast to the case of the full space. A general $v$-th moment of $X$ is considered as a constraint in the principle of maximum Tsallis entropy. The infinite divisibility of the distribution with an arbitrary $v > 0$ and convergence of its $N$-fold convolution to the exact Lévy-stable distribution is discussed in detail. A feature of this derivation is that the Lévy index is related to both the values of $v$ and the index of nonextensivity.

It is known [1,2] that the properties of probability distributions define in the half space are quite different from those in the full space. Physically, such distributions are important in cases where several of the fundamental variables describing systems are bounded from below. Typical examples include time in any irreversible process and the Hamiltonian of a stable system.

In recent years, the Lévy distribution has been invoked to study many physical phenomena, such as anomalous diffusion [3], power-law responses of materials [4], aging in glassy systems [5], and so on. (See Ref. [6] for other topics.) The Lévy-type distribution in the full space has been analyzed based on nonextensive statistics [7] with the Tsallis entropy [8]. The distribution derived therein has the same asymptotic behavior as the Lévy distribution for a large value of the random variable. Combined with the Lévy-Gnedenko generalized central limit theorem [1,2], this is expected to be a reliable procedure for obtaining the Lévy distribution based on maximum entropy principle. Though the Lévy distribution in the half space has been discussed in the literature and has been used in physical contexts [4,9], the consideration of this problem in the light of maximum entropy principle has not been given so far. The purpose of this paper is to provide such a demonstration for the half-space Lévy-type distribution to fill this gap.

The Lévy-stable distribution in the half space $(0, \infty)$ is defined by [2]

$$L_\gamma(x) = \frac{1}{2\pi} \int_{-\infty}^{\infty} dk \, \chi_L(k) \, e^{-ikx}. \qquad (1)$$

In this equation, $\chi_L(k)$ is the characteristic function, which is given by

$$\chi_L(k) = \exp\left\{-a|k|^\gamma \exp\left[i\varepsilon(k)\frac{\theta\pi}{2}\right]\right\}, \qquad (2)$$

where $a$ is a positive constant, $\gamma$ the Lévy index, $\theta$ a constant satisfying $|\theta| \leq \gamma$ and $\varepsilon(k) = k/|k|$ the sign function of $k$. In contrast to the Lévy distribution in the full space whose index is in the range $0 < \gamma < 2$, in the half space the index is confined to

$$0 < \gamma < 1. \tag{3}$$

This distribution has the following asymptotic behavior for large values of $x$:

$$L_\gamma(x) \sim x^{-1-\gamma}, \tag{4}$$

and therefore all of the ordinary positive-integer moments, $\langle X^n \rangle_1 = \int_0^\infty dx\, x^n\, L_\gamma(x)$ ($n = 1, 2, 3, \mathrm{L}$), are divergent. This should be compared with the case in the full space where the (only) ordinary first moment can be finite. (In the notation used here, $x$ is the value of the random variable $X$.) It is known that the Lévy distributions $\{L_\gamma(x): 0 < \gamma < 1\}$ form a stable class in the sense that the sum of $N$ scaled independent random variables $\{x_i\}_{i=1,2,\mathrm{L},N}$

$$x = \frac{x_1 + x_2 + \mathrm{L} + x_N}{B_N} \tag{5}$$

obeys the same distribution as Lévy's. The scaling factor $B_N$ has to be chosen in such a way that the limit distribution is independent of the number $N$ of convolutions. Suppose each $x_i$ obey the Lévy distribution in eq. (1). Then, the distribution of $x$ is given by

$$L_\gamma^{(N)}(x) = B_N \left( L_\gamma * \mathrm{L} * L_\gamma \right)(B_N x), \tag{6}$$

where

$$(f * g)(x) = \int_0^x dx'\, f(x - x')\, g(x'). \tag{7}$$

The characteristic function of $L_\gamma^{(N)}(x)$ is calculated to be

$$\chi_L^{(N)}(k) = \int_0^\infty dx\, e^{ikx}\, L_\gamma^{(N)}(x)$$

$$= \left[ \chi_L \left( \frac{k}{B_N} \right) \right]^N, \qquad (8)$$

In calculating the inverse Fourier transformation, note that $L_\gamma(x) = 0$ for $x < 0$. Thus, we find

$$B_N = N^{1/\gamma}, \qquad (9)$$

where, without loss of generality, we may set the constant of proportionality to be unity.

The derivation of the Lévy distribution in the full space based on the ordinary maximum entropy principle with the Boltzmann-Shannon entropy turned out to be unsatisfactory as shown in Ref. [10]. This difficulty was overcome by using Tsallis' nonextensive generalization of the maximum entropy principle [7] with the constraint on the generalized second moment defined by $\langle X^2 \rangle_q \equiv \int_{-\infty}^{\infty} dx\, x^2 \, [p(x)]^q \Big/ \int_{-\infty}^{\infty} dx\, [p(x)]^q$, where $p(x)$ is the distribution to be determined by this principle and $q$ is the index of nonextensivity. So far, this formulation is given only in the full space. In what follows, we examine the corresponding problem in the half space. In this consideration, to keep the discussion general, we treat a class of the $v$-th moments in place of the generalized second moment as a constraint.

The nonextensive statistics formalism is based on the Tsallis entropy [7,8]

$$S_q[p] = \frac{1}{1-q} \left\{ \int_0^\infty \frac{dx}{\sigma} [\sigma p(x)]^q - 1 \right\}, \qquad (10)$$

where $\sigma$ is a scale parameter and the distribution $p(x)$ is normalized to unity

$$\int_0^\infty dx\, p(x) = 1. \qquad (11)$$

We maximize $S_q[p]$ under the constraints on the normalization in eq. (11) and on the generalized $v$-th moment defind as

$$\left\langle X^{v}\right\rangle_{q} = \frac{\int\limits_{0}^{\infty} dx\, x^{v}\, [p(x)]^{q}}{\int\limits_{0}^{\infty} dx\, [p(x)]^{q}}, \tag{12}$$

where $v$ is an arbitrary positive number. The resulting distribution function is found to be

$$p(x) = \frac{1}{Z_q(\beta)}\left[1-(1-q)\frac{\tilde{\beta}}{c_q}\left(x^{v}-\left\langle X^{v}\right\rangle_{q}\right)\right]^{1/(1-q)}, \tag{13}$$

where $Z_q(\beta)$ is the normalization constant, $\tilde{\beta} = \beta/\sigma^{q-1}$ with $\beta$ the Lagrange multiplier associated with the constraint in eq. (12) and $c_q = \int_0^{\infty} dx\, [p(x)]^q$. In the present work, since we are interested in the Lévy problem, we focus on the distribution with a long tail. Then, $q$ should be larger than unity. The normalizability condition requires $q < v+1$. In addition, the divergence of the ordinary first moment $\langle X \rangle_1$ places another condition $q \geq v/2+1$. Therefore, the range of interest is

$$\frac{v}{2}+1 \leq q < v+1. \tag{14}$$

It should be noted that this condition guarantees the finiteness of $\langle X^{v} \rangle_q$. The quantities $c_q$ and $\langle X^{v} \rangle_q$ have to be calculated by using $p(x)$ self-consistently. The normalization condition on $p(x)$ leads to the identical relation

$$c_q = [Z_q(\beta)]^{1-q}. \tag{15}$$

Let us rewrite eq. (13) in the form

$$p(x) = \frac{1}{Z_q(\beta)}\left(\frac{sc_q}{\tilde{\beta}}\right)^{s}\left(\xi^{v}+x^{v}\right)^{-s}, \tag{16}$$

where

$$s = \frac{1}{q-1}, \tag{17}$$

$$\xi^v = \frac{sc_q}{\tilde{\beta}} - \langle X^v \rangle_q. \tag{18}$$

The range of $q$ in eq. (14) is now written in terms of $s$ as

$$1 < vs < 2. \tag{19}$$

From eqs. (16) and (18), consistency requires that $\xi^v$ be positive. That this turns out to be consistent with the above inequality will be ascertained subsequently. Using the distribution function in eq. (16), we have the following results:

$$Z_q(\beta) = \frac{\xi^{1-vs}}{v}\left(\frac{sc_q}{\tilde{\beta}}\right)^s B\left(\frac{1}{v}, s - \frac{1}{v}\right), \tag{20}$$

$$\langle X^v \rangle_q = \frac{\xi^{1-vs}}{vZ_q(\beta)}\left(\frac{sc_q}{\tilde{\beta}}\right)^{s+1} B\left(\frac{1}{v}+1, s - \frac{1}{v}\right), \tag{21}$$

where $B(w, z)$ is the beta function. Combining these results, we obtain

$$\langle X^v \rangle_q = \frac{c_q}{v\tilde{\beta}}. \tag{22}$$

Substituting this into eq. (18) and taking eq. (19) into account, we confirm that $\xi^v$ is positive as promised above. Finally, the distribution is found to be

$$p(x) = \frac{v\xi^{vs-1}}{B\left(\frac{1}{v}, s - \frac{1}{v}\right)} (\xi^v + x^v)^{-s}. \tag{23}$$

Now, we are in a position to examine the properties of convergence of this distribution to the Lévy-stable distribution after *N*-fold convolution for large *N*. The characteristic function of the distribution of $x$ in eq. (5) is

$$\chi^{(N)}(k) = \int_0^\infty dx\, e^{ikx}\, B_N (p * p * \text{L} * p)(B_N x)$$

$$= \left[\chi\left(\frac{k}{B_N}\right)\right]^N, \tag{24}$$

where $\chi(k)$ is the characteristic function of $p(x)$. Following Ref. [1], we express $\chi(k/B_N)$ as follows:

$$\chi\left(\frac{k}{B_N}\right) = 1 + \int_0^\infty dx \left[\exp\left(\frac{i\varepsilon(k)|k|x}{B_N}\right) - 1\right] p(x)$$

$$= 1 + \frac{v\xi^{vs-1}}{B\left(\frac{1}{v}, s - \frac{1}{v}\right)} \left(\frac{|k|}{B_N}\right)^{vs-1} \int_0^\infty dy\, \frac{\exp[i\varepsilon(k)y] - 1}{y^{vs}}$$

$$+ \frac{v\xi^{vs-1}}{B\left(\frac{1}{v}, s - \frac{1}{v}\right)} \left(\frac{|k|}{B_N}\right)^{vs-1}$$

$$\times \int_0^\infty dy\, \{\exp[i\varepsilon(k)y] - 1\} \left\{\frac{1}{\left[y^v + \left(\frac{\xi|k|}{B_N}\right)^v\right]^s} - \frac{1}{y^{vs}}\right\}. \tag{25}$$

To evaluate the integral, we use the following result given in Ref. [1] (p. 169):

$$\int_0^\infty dy \frac{e^{\pm iy} - 1}{y^{1+\alpha}} = -\exp\left(mi\frac{\alpha\pi}{2}\right)|L(\alpha)|, \quad (26)$$

where $L(\alpha)$ is given by

$$L(\alpha) = \int_0^\infty \frac{dy}{y^{1+\alpha}}\left(e^{-y} - 1\right) < 0 \quad (0 < \alpha < 1). \quad (27)$$

Equation (26) is then calculated to be

$$\chi\left(\frac{k}{B_N}\right) = 1 - a\left(\frac{|k|}{B_N}\right)^{vs-1} \exp\left[i\varepsilon(k)\frac{\theta\pi}{2}\right] + C_N, \quad (28)$$

where

$$a = \frac{v\xi^{vs-1}|L(vs-1)|}{B\left(\frac{1}{v}, s - \frac{1}{v}\right)} > 0, \quad (29)$$

$$\theta = 1 - vs. \quad (30)$$

The term $C_N$ in eq. (28) corresponds to the thrid term in eq. (25) and is the higher-order correction, which vanishes faster than the second term in the limit $N \to \infty$. Taking the logarithm of $\chi^{(N)}(k)$ in eq. (24) and comparing it with eq. (2), we have in the limit $N \to \infty$

$$\ln \chi^{(N)}(k) = N \ln \chi\left(\frac{k}{B_N}\right)$$

$$= -a|k|^\gamma \exp\left[i\varepsilon(k)\frac{\theta\pi}{2}\right], \quad (31)$$

where the Lévy index $\gamma$ is identified to be

$$\gamma = \nu s - 1, \tag{32}$$

and $B_N$ is the same as that given in eq. (9). From eq. (19), equation (32) is seen to be consistent with eq. (3). Equation (31) shows that the nonextensive statistical distribution in the half space is infinitely divisible for an arbitrary $\nu > 0$ and approaches the Lévy-stable distribution by the $N$-fold convolution for large $N$. Equation (32) reveals an important feature of the approach presented here in that the Lévy index is related to both the order $\nu$ of the moment and the index of nonextensivity $q = 1 + 1/s$.

In conclusion, we have examined the nonextensive statistical formalism to investigate the Lévy distribution in the half space. We have considered the generalized $\nu$-th moment of the basic random variable and have found the self-consistent solution of the nonextensive statistical distribution. We have shown that this distribution is infinitely divisible for an arbitrary $\nu > 0$ and, in full conformity with the generalized central limit theorem, the $N$-fold convolution of the distribution converges to the Lévy-stable distribution in the large-$N$ limit.

One of us (A. K. R.) acknowledges the partial support from the US Office of Naval Research. The other (S.A.) was supported by the GAKUJUTSU-SHO Program of College of Science and Technology, Nihon University. He thanks the warm hospitality of the Naval Research Laboratory, Washington, DC, which made this collaboration possible.